\documentclass[letterpaper,10pt,conference]{IEEEtran}
\IEEEoverridecommandlockouts 

\makeatletter
\newcommand{\AddInputPath}[1]{%
  \ifx\input@path\@undefined
    \def\input@path{#1}
  \else
    \g@addto@macro{\input@path}{#1}
  \fi
}
\makeatother

\usepackage{shellesc}

\usepackage{etex}

\usepackage[font=footnotesize,caption=false]{subfig} 

\usepackage{relsize}

\usepackage[dvipsnames,svgnames,usenames]{xcolor}
\usepackage{array}
\usepackage{booktabs,tabularx}
\usepackage{multirow}
\usepackage[binary-units=true,per-mode=symbol,detect-mode=true]{siunitx}
\usepackage{graphicx}

\graphicspath{{images/}} 

\usepackage[T1]{fontenc}
\usepackage{textcomp}
\usepackage[utf8]{inputenc}
\usepackage[final]{microtype}
\usepackage{icomma}
\usepackage{xspace}

\usepackage[tbtags]{amsmath}
\usepackage{amssymb,amsfonts,bm}
\usepackage{mathtools} 
\usepackage{dsfont}
\usepackage{mathrsfs}
\usepackage{accents}
\usepackage{empheq}
\usepackage{nccmath}
\usepackage{balance}
\usepackage{setspace}

\usepackage{color}
\usepackage{calc}
\usepackage{tikz}
\usepackage{pgfplots,pgfplotstable}

\usepackage{pdftexcmds}
\makeatletter
\newcommand{\strequal}[2]{\pdf@strcmp{#1}{#2}==0}
\makeatother

\usepackage[final,bookmarksnumbered=true,pdfborder={0 0 0}]{hyperref}
\usepackage[capitalize]{cleveref}
\usepackage{refcount}

\usepackage[inline]{enumitem}
\usepackage{algorithm}
\usepackage{algpseudocode}
\makeatletter
\newcommand{\algmargin}{\the\ALG@thistlm}
\algblockx{StartIA}{EndIA}[1]{\textbf{Input} #1}[1]{\textbf{Return} #1}
\makeatother
\newlength{\whilewidth}
\algdef{SE}[parWHILE]{parWhile}{EndparWhile}[1]
  {\parbox[t]{\dimexpr\linewidth-\algmargin}{%
     \hangindent\whilewidth\strut\algorithmicwhile\ #1\ \algorithmicdo\strut}}{\algorithmicend\ \algorithmicwhile}%
\algnewcommand{\parState}[1]{\State%
  \parbox[t]{\dimexpr\linewidth-\algmargin}{\strut #1\strut}}

\makeatletter
\newcommand\fs@spaceruled{\def\@fs@cfont{\bfseries}\let\@fs@capt\floatc@ruled
  \def\@fs@pre{\vspace{.05in}\hrule height.8pt depth0pt \kern2pt}%
  \def\@fs@post{\kern2pt\hrule\relax}%
  \def\@fs@mid{\kern2pt\hrule\kern2pt}%
  \let\@fs@iftopcapt\iftrue}
\makeatother

\floatstyle{spaceruled}
\restylefloat{algorithm}

\usepackage{glossaries-prefix}
\usepackage{ifthen}
\usepackage[noadjust]{cite}
\usepackage{multibib}

\usepackage{comment}
\usepackage{todonotes}
\let\legacytodo\todo
\newcommand{\ruggedtodo}[2][]{\tikzexternaldisable\legacytodo[#1]{#2}\tikzexternalenable}
\renewcommand{\todo}[1]{\ruggedtodo[inline]{#1}}

\makeglossaries

\newacronym{soa}{SoA}{state-of-the-art}
\newacronym{mh}{MH}{multi-hop}
\newacronym{tcs}{TCS}{time correlated sparsification}
\newacronym{ia}{IA}{incremental aggregation}
\newacronym{sfl}{SFL}{satellite federated learning}
\newacronym{esa}{ESA}{European Space Agency}
\newacronym{tle}{TLE}{two-line element set}
\newacronym{ai}{AI}{artificial intelligence}
\newacronym{ann}{ANN}{artificial neural network}
\newacronym{jscc}{JSCC}{joint source-channel coding}
\newacronym{raan}{RAAN}{right ascension of the ascending node}
\newacronym{uav}{UAV}{unmanned aerial vehicle}
\newacronym{haps}{HAPS}{high-altitude platform station}
\newacronym{6g}{6G}{sixth generation}
\newacronym{cgr}{CGR}{contact graph routing}
\newacronym{dtn}{DTN}{delay-tolerant networking}
\newacronym{fl}{FL}{federated learning}
\newacronym{dl}{DL}{deep learning}
\newacronym{fedavg}{FedAvg}{federated averaging}
\newacronym{dml}{DML}{distributed ML}
\newacronym{ps}{PS}{parameter server}
\newacronym[prefix={an\space},prefixfirst={a~}]{ml}{ML}{machine learning}
\newacronym{sgd}{SGD}{stochastic gradient descent}
\newacronym{dsgd}{DSGD}{distributed stochastic gradient descent}
\newacronym{isl}{ISL}{inter-satellite link}
\newacronym{gsl}{GSL}{ground to satellite link}
\newacronym{gs}{GS}{ground station}
\newacronym{ecef}{ECEF}{earth-centered, earth-fixed}
\newacronym{eci}{ECI}{Earth-centered inertial}
\newacronym{ofdm}{OFDM}{orthogonal frequency-division multiplexing}
\newacronym{cp}{CP}{cyclic prefix}
\newacronym{los}{LOS}{line-of-sight}
\newacronym{leo}{LEO}{low earth orbit}
\newacronym{meo}{MEO}{medium earth orbit}
\newacronym{gso}{GSO}{geosynchronous orbit}
\newacronym{geo}{GEO}{geostationary}
\newacronym{eo}{EO}{Earth observation}
\newacronym{iot}{IoT}{Internet of Things}
\newacronym{irs}{IRS}{intelligent reflecting surface}
\newacronym{socp}{SOCP}{second-order cone program}
\newacronym{soc}{SOC}{second-order cone}
\newacronym{dsl}{DSL}{digital subscriber line}
\newacronym{wsee}{WSEE}{weighted sum energy efficiency}
\newacronym{mmwave}{mmWave}{millimeter wave}
\newacronym{dfg}{DFG}{Deutsche Forschungsgemeinschaft}
\newacronym{haec}{HAEC}{Highly Adaptive Energy-Efficient Computing}
\newacronym{hpc}{HPC}{High Performance Computing}
\newacronym{mac}{MAC}{multiple-access channel}
\newacronym{bc}{BC}{broadcast channel}
\newacronym{siso}{SISO}{single-input single-output}
\newacronym{simo}{SIMO}{single-input multiple-output}
\newacronym{miso}{MISO}{multiple-input single-output}
\newacronym{mimo}{MIMO}{multiple-input multiple-output}
\newacronym{af}{AF}{amplify-and-forward}
\newacronym{df}{DF}{decode-and-forward}
\newacronym{cf}{CF}{compress-and-forward}
\newacronym{mwrc}{MWRC}{multi-way relay channel}
\newacronym{dmmwrc}{DM-MWRC}{discrete memoryless multi-way relay channel}
\newacronym{pde}{PDE}{partial data exchange}
\newacronym{fde}{FDE}{full data exchange}
\newacronym{iid}{i.i.d.\@}{independent and identically distributed}
\newacronym{di}{DI} {difference of increasing}
\newacronym{dc}{DC}{difference of convex}
\newacronym{mm}{MM}{mixed monotonic}
\newacronym{mmp}{MMP}{mixed monotonic programming}
\newacronym{awgn}{AWGN}{additive white Gaussian noise}
\newacronym{wgn}{WGN}{white Gaussian noise}
\newacronym{awg}{AWG}{additive white Gaussian}
\newacronym{sic}{SIC}{successive interference cancellation}
\newacronym{snr}{SNR}{signal-to-noise ratio}
\newacronym{sinr}{SINR}{signal to interference plus noise ratio}
\newacronym{inr}{INR}{interference to noise ratio}
\newacronym{zf}{ZF}{zero-forcing}
\newacronym{mrt}{MRT}{maximum ratio transmission}
\newacronym{mmse}{MMSE}{minimum mean square error}
\newacronym{sud}{SUD}{single user decoding}
\newacronym{dof}{DoF}{degrees of freedom}
\newacronym{gdof}{GDoF}{generalized degrees of freedom}
\newacronym{nnc}{NNC}{noisy network coding}
\newacronym{dmn}{DMN}{discrete memoryless network}
\newacronym{csi}{CSI}{channel state information}
\newacronym{pmf}{pmf}{probability mass function}
\newacronym{dmic}{DM-IC}{discrete memoryless interference channel}
\newacronym{ic}{IC}{interference channel}
\newacronym{gic}{GIC}{Gaussian interference channel}
\newacronym{if}{IF}{interference}
\newacronym{ee}{EE}{energy efficiency}
\newacronym{gee}{GEE}{global energy efficiency}
\newacronym{tin}{TIN}{treating interference as noise}
\newacronym{snd}{SND}{simultaneous non-unique decoding}
\newacronym{sd}{SD}{simultaneous decoding}
\newacronym{hk}{HK}{Han-Kobayashi}
\newacronym{rs}{RS}{rate splitting}
\newacronym{rf}{RF}{radio frequency}
\newacronym{pa}{PA}{power amplifier}
\newacronym{lna}{LNA}{low noise amplifier}
\newacronym{lo}{LO}{local oscillator}
\newacronym{adc}{ADC}{analog-to-digital converter}
\newacronym{dac}{DAC}{digital-to-analog converter}
\newacronym{dsp}{DSP}{digital signal processing}
\newacronym{brd}{BRD}{best response dynamics}
\newacronym{br}{BR}{best response}
\newacronym{ne}{NE}{Nash equilibrium}
\newacronym{lhs}{LHS}{left-hand side}
\newacronym{rhs}{RHS}{right-hand side}
\newacronym{ran}{RAN}{radio access network}
\newacronym{qos}{QoS}{Quality of Service}
\newacronym{ngmn}{NGMN}{Next Generation Mobile Networks}
\newacronym{cap}{CAP}{Capacity Adaptation}
\newacronym{bwa}{BW}{Bandwidth Adaptation}
\newacronym{prb}{PRB}{physical resource block}
\newacronym{se}{SE}{spectral efficiency}
\newacronym{tp}{TP}{throughput}
\newacronym{bs}{BS}{base station}
\newacronym{ue}{UE}{user equipment}
\newacronym{mop}{MOP}{multi-objective optimization problem}
\newacronym{gda}{GDA}{generalized Dinkelbach's algorithm}
\newacronym{midcp}{MIDCP}{mixed integer disciplined convex programming}
\newacronym{lp}{LP}{linear program}
\newacronym{brb}{BRB}{branch reduce and bound}
\newacronym{bb}{BB}{branch and bound}
\newacronym{sit}{SIT}{successive incumbent transcending}
\newacronym{oma}{OMA}{orthogonal multiple access}
\newacronym{noma}{NOMA}{non-orthogonal multiple access}
\newacronym{wlog}{w.l.o.g.\@}{without loss of generality}
\newacronym{lsc}{l.s.c.\@}{lower semi-continuous}
\newacronym{usc}{u.s.c.\@}{upper semi-continuous}
\newacronym{kkt}{KKT}{Karush-Kuhn-Tucker}
\newacronym{ptp}{PTP}{point-to-point}

\glsenableentrycount
\makeglossaries

\usetikzlibrary{positioning}
\usetikzlibrary{calc}
\usetikzlibrary{math}
\usetikzlibrary{fit}
\usetikzlibrary{intersections}
\usetikzlibrary{decorations.pathreplacing}
\usetikzlibrary{decorations.markings}
\usetikzlibrary{3d,angles}
\usetikzlibrary{arrows.meta}

\pgfdeclarelayer{background}
\pgfsetlayers{background,main}

\usetikzlibrary{external}

\usepgfplotslibrary{colorbrewer}

\tikzset{
	small1/.style={fill=DeepPink},
	small2/.style={fill=DeepSkyBlue},
	small3/.style={fill=MediumSpringGreen},
	ps/.style={fill=Gold},
	link/.style = {semithick},
	plane/.style={plane origin={(#1,0,0)}, plane x = {(#1,0,1)}, plane y = {(#1,1,0)}, rotate around y = -9, canvas is plane}
}

\tikzset{
	antenna/.pic={
		\draw[thick] (0,0) -- ++(120:2mm) -- ++(0:2mm) -- cycle -- (0,-1.5mm);
	}
}


\crefname{equation}{}{}
\crefrangeformat{equation}{(#3#1#4)--(#5#2#6)}
\crefmultiformat{equation}{(#2#1#3)}{ and~(#2#1#3)}{, (#2#1#3)}{, (#2#1#3)}
\crefrangemultiformat{equation}{#3(#1)#4--#5(#2)#6}{, #3(#1)#4--#5(#2)#6}{, #3(#1)#4--#5(#2)#6}{, #3(#1)#4--#5(#2)#6}
\crefrangeformat{algorithm}{Algorithms~#3#1#4--#5#2#6}

\undef\mod
\DeclareMathOperator\mod{mod}

\let\vec\bm

\allowdisplaybreaks[3]

\DeclareSIUnit \dBm {dBm}
\DeclareSIUnit \dBW {dBW}
\DeclareSIUnit \bpcu {bpcu}

\DeclareFontFamily{U}{mathx}{\hyphenchar\font45}
\DeclareFontShape{U}{mathx}{m}{n}{
      <5> <6> <7> <8> <9> <10>
      <10.95> <12> <14.4> <17.28> <20.74> <24.88>
      mathx10
      }{}
\DeclareSymbolFont{mathx}{U}{mathx}{m}{n}
\DeclareMathSymbol{\bigtimes}{1}{mathx}{"91}

\newtheorem{proposition}{Proposition}

\pgfplotscreateplotcyclelist{default}{%
	blue,mark=*\\%
	red,mark=star\\%
	teal,mark=square*\\%
	brown!60!black,mark=otimes*\\%
}

\definecolor{plot1}{RGB}{228,26,28}
\definecolor{plot2}{RGB}{55,126,184}
\definecolor{plot3}{RGB}{77,175,74}
\definecolor{plot4}{RGB}{152,78,163}
\definecolor{plot5}{RGB}{255,127,0}
\definecolor{plot6}{RGB}{166,86,40}
\tikzstyle{fedsatschedule}=[plot1]
\tikzstyle{fedsat}=[plot2]
\tikzstyle{fedisl}=[plot3]
\tikzstyle{fedavg}=[plot4]
\tikzstyle{fedasync1}=[plot5]
\tikzstyle{fedasync2}=[plot6]

\newcolumntype{P}[1]{>{\centering\arraybackslash}p{#1}}

\ifCLASSOPTIONdraftcls
\AtBeginEnvironment{figure}{}
\fi

\AtBeginEnvironment{algorithmic}{\footnotesize}

\begin{document}
\bstctlcite{IEEEexample:BSTcontrol}
\title{Sparse Incremental Aggregation in\\ Multi-Hop Federated Learning}

\author{\IEEEauthorblockN{Sourav Mukherjee\IEEEauthorrefmark{1}, Nasrin Razmi\IEEEauthorrefmark{1}, Armin Dekorsy\IEEEauthorrefmark{1}, Petar Popovski\IEEEauthorrefmark{2}\IEEEauthorrefmark{1}, Bho Matthiesen\IEEEauthorrefmark{1}}
	\IEEEauthorblockA{\IEEEauthorrefmark{1}University of Bremen, Department of Communications Engineering, Germany\\\IEEEauthorrefmark{2}Aalborg University, Department of Electronic Systems, Denmark\\ email: \{mukherjee, razmi, dekorsy, matthiesen\}@ant.uni-bremen.de, petarp@es.aau.dk}
\thanks{
This work is supported by the German Research Foundation (DFG) under Grant EXC 2077 (University Allowance).
}%
}
\maketitle

\begin{abstract}
This paper investigates federated learning (FL) in a 	
multi-hop communication setup, such as in constellations with inter-satellite links. In this setup, part of the FL clients are responsible for forwarding other client's results to the parameter server. Instead of using conventional routing, the communication efficiency can be improved significantly by using in-network model aggregation at each intermediate hop, known as \emph{incremental aggregation (IA)}. Prior works \cite{Razmi2024} have indicated diminishing gains for IA under gradient sparsification. Here we study this issue and propose several novel correlated sparsification methods for IA. Numerical results show that, for some of these algorithms, the full potential of IA is still available under sparsification without impairing convergence. We demonstrate a 15$\times$ improvement in communication efficiency over conventional routing and a 11$\times$ improvement over \cgls{soa} sparse IA.
\end{abstract}
\glsresetall

\section{Introduction}
\cGls{fl} \cite{mcmahan2017communication} is an instance of distributed \cgls{ml} over bandwidth-restricted communication networks, in which a large number of clients use the local sets and collaboratively train \pgls{ml} model. This process is orchestrated by a central \cgls{ps}, which is responsible for synchronizing intermediate results among the clients. This involves collecting intermediate results $\{\vec g_k^t \}_k$ from all clients, aggregating these values into a new global iterate $\vec w^{t+1}$ of the \cgls{ml} model parameters, and distributing $\vec w^{t+1}$ back to all clients. As contemporary \cgls{ml} models can consist of billions of parameters \cite{bengio2016}, communication efficiency in this synchronization phase is paramount.

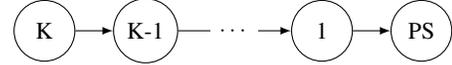
\begin{figure}
	\centering
	\begin{tikzpicture}[every path/.style={-latex}, font=\footnotesize]
		\begin{scope}[every node/.style = {draw,circle,minimum width=2.3em, font=\small}, node distance = .5cm]
			\node (K) {K};
			\node[right=of K] (K-1) {K-1};
			\node[right=1.5cm of K-1] (1) {1};
			\node[right=of 1] (ps) {PS};
		\end{scope}

		\draw (K) -- (K-1);
		\draw (1) -- (ps);

		\draw (K-1) --node[fill=white] {$\cdots$} (1);
	\end{tikzpicture}
	\caption{Multi-hop federated learning system.}\label{fig:sysmod}
\end{figure}

Here, we are interested in \gls{fl} over \cgls{mh} networks, where some of the \cgls{fl} clients are responsible for forwarding other client's results to the \cgls{ps}.
 Such a setup arises in the implementation of \cgls{fl} in satellite constellations \cite{network2022,Razmi2022}, when inter-satellite links are used for communicating with the \cgls{ps} \cite{Razmi2022a,Razmi2024}. Similar setups are also applicable to \cgls{fl} in wireless mesh and multi-hop sensor networks \cite{Chen2022b}.
It is shown in \cite{Razmi2022a,Razmi2024,Chen2022b} that a network topology-aware implementation of \cgls{fl}, leveraging in-network computing,
leads to a massive reduction in the communication load during the result-collection phase.
To elaborate, consider the \cgls{mh} network in \cref{fig:sysmod}, where nodes $1, \dots, K$ are \cgls{fl} clients. Using conventional routing to transmit $\{\vec g_k^t \}_k$ results in a total of
$1 + 2 + \dots + (K-1) + K = (K^2 + K)/2$
transmissions the size of $\vec g_k^t$. However, the \cgls{ps} is primarily interested in the weighted sum of $\{\vec g_k^t \}_k$ to compute the new model iterate. By computing this weighted sum incrementally within the network, i.e., each hop combines their own update with the results of previous hops before forwarding, each node needs only transmit a single vector of size $\vec g_k^t$, resulting in $K$ transmissions in total.
However, it has been observed in \cite{Razmi2024} that the efficiency of this \cgls{ia} procedure is reduced massively when combined with gradient sparsification methods such as Top-$Q$ \cite{Aji2017,Alistarh2018}. This is because individual gradient sparsification at each node leads to (almost) uncorrelated sparsification supports. The result is that the number of non-zero elements in the partial aggregate increases with each hop.

In this paper, we consider sparse \cgls{ia} and propose several approaches to the issues outlined above. After establishing the system model in \cref{sec:sysmod}, we analyze the shortcomings of \cgls{soa} sparse \cgls{ia} and take two different angles towards improving it in \cref{sec:algs}. Both methods rely on intentionally creating correlation in the sparsification procedure. In \cref{sec:tcs}, we combine our methods with \cgls{tcs} \cite{Ozfatura2021a} to further increase this correlation. The communication cost of the proposed algorithms is discussed in \cref{sec:cost}, and evaluated numerically in \cref{sec:numeval}.

\paragraph*{Notation} 
The $L^2$ vector norm is $\Vert \vec x\Vert$, $\Vert \vec x\Vert_0$ is the number of nonzero elements in $\vec x$, and $\mathds 1(\vec x)$ is the indicator vector of $\vec x$. The Hadamard product of two vectors $\vec a, \vec b$ is $\vec a \circ \vec b$. The operations $\lceil \cdot \rceil$ and $\lfloor \cdot \rceil$ round to the nearest greater and nearest integer, respectively. The function $S(\vec x, Q)$ returns the Top-$Q$ sparsification of $\vec x$ and $s(\vec x, Q)= \mathds 1(S(\vec x, Q))$ returns the corresponding sparsification mask.

\section{System Model} \label{sec:sysmod}
Consider the \cgls{mh} communication system in \cref{fig:sysmod} with $K+1$ nodes, where node $k$, $k = 1, \dots, K-1$, is connected to nodes $k-1$ and $k+1$, node $K$ connects to node $K-1$, and node 0, denoted as \acrshort{ps}, is connected to node 1.
These nodes form a \cgls{fl} system in which nodes $1, \dots, K$ collaborate to solve \pgls{ml} training problem
$\min_{\vec w \in\mathds R^d} \frac{1}{D} \sum_{k=1}^K \sum_{\vec x\in\mathcal D_k} f(\vec x; \vec w)$,
with per-sample loss function $f(\vec x; \vec w)$, $d$-dimensional model parameter vector $\vec w$, and $\mathcal D_k$ the data set of client $k$. We define $D_k = |\mathcal D_k |$ and the total number of samples $D = \sum_k D_k$. Nodes $1, \dots, K$ are also referred to as clients.

Local data sets $\mathcal D_k$ are not shared among clients, making a distributed solution of the training problem necessary. This is an iterative procedure orchestrated by the \cgls{ps}. In iteration $t$, this \cgls{ps} distributes 
the current model parameter vector $\vec w^t$ to all clients. Then, each client $k$ computes a local update $\vec w^t_k$ to $\vec w^t$, e.g., by performing one or several steps of \cgls{sgd}, and transmits the result to the \cgls{ps}. This is done in the form of the effective gradient $\vec g_k^t = \vec w_k^t - \vec w^t$.
After collecting all updates $\{ \vec g^t_k \}_k$, the \cgls{ps} computes 
a new iteration of the model parameters as \vspace{-2pt}
\begin{equation} \label{eq:modupd}
	\vec w^{t+1} = \vec w^t + \frac{1}{D} \sum\nolimits_{k = 1}^K D_k \vec g_{k}^t.
\end{equation} 
This process is repeated until convergence.
We are primarily interested in the aggregation phase. To this end, we are assuming that the effective gradients $\{ \vec g_k^t \}_k$ are obtained by \emph{some} means at the clients and that the \cgls{ps} is only interested in their aggregate value $\sum_{k = 1}^K D_k \vec g_{k}^t$.
\vspace{-5pt}
\subsection{Sparse Incremental Aggregation}
Observe from \cref{eq:modupd} that the \cgls{ps} is not interested in individual updates but only in the weighted sum $\sum_{k = 1}^K D_k \vec g_{k}^t$. This is exploited in \cgls{ia} as follows. Instead of forwarding its own gradient $\vec g_k^t$ and the previous hop's gradients $\{ \vec g_j^t \}_{j=k-1}^K$ separately, node $k$ waits for all previous nodes and computes the partial aggregate 
$\vec\gamma_k^t = \sum_{i = k}^K D_i \vec g_i^t$.
This is then transmitted to the PS via the next hop. Applying this approach in each node $k$, $k = 1, \dots, K-1$, we observe that node $k$ will only receive a single transmission from node $k+1$ with the partial aggregate $\vec\gamma_{k+1}^t$. Then, it computes its own partial aggregate as
\begin{equation} \label{eq:incagg}
	\vec\gamma_k^t = \vec\gamma_{k+1}^t + D_k \vec g_k^t
\end{equation}
and forwards it to node $k-1$. Thus, the PS receives $\vec\gamma_1^t = \sum_{k=1}^K D_k \vec g_k^t$ and computes $\vec w^{t+1} = \vec w^t + \frac{1}{D} \vec\gamma_1^t$.

Our goal is to combine \cgls{ia} with Top-$Q$ gradient sparsification \cite{Aji2017,Alistarh2018}. The Top-$Q$ procedure sets all gradient entries except the $Q$ largest magnitude values to zero and transmits the resulting vector in sparse representation. This results in a significant bandwidth reduction, as only the nonzero entries (and their indices) need to be transmitted. Top-$Q$ sparsification is commonly implemented with an error feedback mechanism to improve convergence. In particular, let $\vec e_k^{t-1}$ be node $k$'s sparsification error from the last iteration. Then, the  error-compensated effective gradient at node $k$ is $\tilde{\vec g}_k^t = \vec g_k^t + \vec e_k^{t-1}$. Based on this vector and the incoming $\vec\gamma_{k+1}^t$, a new sparse partial aggregate $\vec\gamma_k^t$ and error vector $\vec e_k^t$ are computed.

In this paper, we explore several methods for computing $\vec\gamma_k$ from $\tilde{\vec g}_k^t$ and $\vec\gamma_{k+1}^t$. The \cgls{soa} approach \cite{Razmi2024} is a direct concatenation of \cgls{ia} and Top-$Q$ sparsification applied to $\tilde{\vec g}_k^t$. That is,  node $k$ computes $\bar{\vec g}_k^t = S(\tilde{\vec g}_k^t, Q)$ and $\vec e^t_k = \tilde{\vec g}_k^t - \bar{\vec g}_k^t$, where $S(\cdot, Q)$ is the Top-$Q$ sparsification operation. The outgoing partial aggregate $\vec\gamma_k^t$ is then computed as $\vec\gamma_{k+1}^t + D_k \bar{\vec g}_k^t$.
This results in \cref{alg:vanilla} from \cite{Razmi2024}, where lines~\ref{alg:vanilla:1}--\ref{alg:vanilla:3} perform Top-$Q$ sparsification and line~\ref{alg:vanilla:4} implements \cgls{ia}.
Note that scaling by $D_k$ is, equivalently, done in line~\ref{alg:vanilla:1} for consistency with later algorithms.

   \begin{algorithm}
	\caption{Sparse incremental aggregation at node $k$ \cite{Razmi2024}} \label{alg:vanilla}
	\begin{algorithmic}[1]
		\StartIA{$\vec g_k^t$, $\vec\gamma_{k+1}^t$}
			\State Error feedback $\tilde{\vec g}_k^t \gets D_k \vec g_k^t + \vec e_k^{t-1}$ \label{alg:vanilla:1}
			\State Sparsification $\bar{\vec g}_k^t \gets S(\tilde g_k^t, Q)$ \label{alg:vanilla:2}
			\State Update error $\vec e_k^t \gets \tilde{\vec g}_k^t - \bar{\vec g}_k^t$ \label{alg:vanilla:3}
			\State Incremental Aggregation $\vec\gamma_k^t \gets \bar{\vec g}_k^t + \vec\gamma_{k+1}^t$ \label{alg:vanilla:4}
		\EndIA{$\vec\gamma_k^t$}
	\end{algorithmic}
\end{algorithm}

\section{Sparse Incremental Aggregation, Revisited} \label{sec:algs}
Consider the sparse \cgls{ia} operation in Line~\ref{alg:vanilla:4} of \cref{alg:vanilla}. If $\bar{\vec g}_k^t$ and $\vec\gamma_{k+1}^t$ have their nonzero elements in exactly the same positions, i.e., they have the same sparsification support, the outgoing partial aggregate $\vec\gamma_k$ will have exactly $Q$ nonzero elements. This, however, is usually not the case and the number of nonzero elements in $\vec\gamma_k^t$ is
$\max\{ Q,\ \Vert \vec\gamma_{k+1}^t \Vert_0\} \le \Vert \vec\gamma_{k}^t \Vert_0 \le Q + \Vert \vec\gamma_{k+1}^t \Vert_0$,
with a strong tendency towards the upper bound  as $K$ increases and $Q$ decreases \cite{Razmi2024}. Indeed, the results in \cite{Razmi2024} indicate that, as $K \to \infty$, the gain of \cgls{ia} over conventional multiple unicast transmissions completely vanishes for $Q < d$.

\subsection{An Error Minimization Perspective on Sparse IA}
To gain additional insight into this problem, consider a conventional \cgls{fl} setup with direct client-\cgls{ps} links.
Gradient compression, which includes sparsification, is applied to conserve bandwidth in the client-\cgls{ps} link. For any compressor $C(\vec x)$ and node $k$, the compression error is $\Vert \tilde{\vec g}_k^t - C(\tilde{\vec g}_k^t) \Vert^2$ and a compressor is considered optimal if it minimizes this error. Restricting the choice of compressors to the set of sparsification functions $\mathcal S$, it is well established that Top-$Q$ sparsification is optimal under a strict communication budget of transmitting at most $Q$ nonzero elements per iteration \cite[Lemma 2]{Sahu2021}. That is, the optimization problem
\begin{equation} \label{opt:topq}
	\min_{C\in\mathcal S}\enskip \Vert \tilde{\vec g}_k^t - C(\tilde{\vec g}_k^t) \Vert^2 \quad\mathrm{s.t.}\quad \Vert C(\tilde{\vec g}_k^t) \Vert_0 \le Q,
\end{equation}
is solved by $C(\vec x) = S(\vec x, Q)$.

Returning to \cgls{mh} aggregation and \cref{alg:vanilla}, we can make two observations: 1) The outgoing transmission is $\vec\gamma_k^t$ with relevant compression error $\Vert \vec\gamma_k^t - C(\vec\gamma_k^t) \Vert^2$; and 2) the effective transmissions budget is $\widetilde Q = \Vert \vec\gamma_{k+1}^t + S(\tilde{\vec g}_k^t, Q)\Vert_0 \ge Q$. Thus, the error minimization problem corresponding to \cref{opt:topq} is
\begin{equation} \label{opt:mh}
	\min_{C\in\mathcal S}\ \Vert \vec\gamma_k^t - C(\vec\gamma_{k+1}^t, \tilde{\vec g}_k^t) \Vert^2 \enskip\mathrm{s.t.}\enskip \Vert C(\vec\gamma_{k+1}^t, \tilde{\vec g}_k^t) \Vert_0 \le \widetilde Q.
\end{equation}
It is straightforward to show that \cref{alg:vanilla} is strictly suboptimal with respect to \cref{opt:mh}.
\begin{proposition} \label{prop:topq}
	$C(\vec\gamma_{k+1}^t, \tilde{\vec g}_k^t) = S(\tilde{\vec g}_k^t, Q)$ is strictly suboptimal with respect to \cref{opt:mh} unless the sparsification supports of $\vec\gamma_{k+1}^t$ and $\tilde{\vec g}_k^t$ are identical.
\end{proposition}
\begin{IEEEproof}
	Let $\vec m_k^t = s(\tilde{\vec g}_k^t, Q) = \mathds 1(S(\vec x, Q))$ be the Top-$Q$ sparsification mask and $\widetilde{\vec m}_{k+1}^t = \mathds 1(\vec\gamma_{k+1}^t)$ the mask corresponding to the sparsification support of $\vec\gamma_{k+1}^t$, where $\mathds 1(\vec x)$ is the indicator vector of $\vec x$. Consider $C(\vec\gamma_{k+1}^t, \tilde{\vec g}_k^t) = \vec\gamma_{k+1}^t + \mathds 1(\vec m_k^t + \widetilde{\vec m}_{k+1}^t) \circ \tilde{\vec g}_k^t$. Then, the objective of \cref{opt:mh} satisfies
	\begin{align} \label{eq:topqerror}
		\begin{split}
			&\left\Vert \left(\vec 1 - \mathds 1(\vec m_k^t + \widetilde{\vec m}_{k+1}^t)\right) \circ \tilde{\vec g}_k^t \right\Vert^2
			\\{}={}& \sum\nolimits_{i \not\in \mathcal I(\vec m_k^t) \cup \mathcal I(\widetilde{\vec m}_{k+1}^t)} |g_{k,i}^t |^2 \le \sum\nolimits_{i \not\in \mathcal I(\vec m_k^t)} |g_{k,i}^t |^2,
		\end{split}
	\end{align}
	where $\vec 1$ is the all ones vector, $g_{k,i}^t$ is the $i$th element of $\vec g_{k}^t$, and $\mathcal I(\vec x)$ is a set containing the indices of nonzero elements in $\vec x$. The inequality in \cref{eq:topqerror} is strict unless $\sum_{i \not\in \mathcal I(\widetilde{\vec m}_{k+1}^t)\setminus \mathcal I(\vec m_k^t) } |g_{k,i}^t |^2 = 0$, which is the case either if the corresponding elements in $\vec g_k^t$ are zero or $\widetilde{\vec m}_{k+1}^t = \vec m_k^t$.
\end{IEEEproof}
\vspace{-5pt}
\subsection{Reduced-Error Sparse Incremental Aggregation}

The proof of \cref{prop:topq} suggests a direct way to improve \cref{alg:vanilla} without additional communication cost. Instead of transmitting only the nonzero elements in $\vec g_k^t$ returned by Top-$Q$ sparsification, node $k$ can transmit additional gradient elements within the nonzero positions of $\vec\gamma_{k+1}^t$. This is described in \cref{alg:reducederror}, where the sparsification in line~\ref{alg:vanilla:2} of \cref{alg:vanilla} is replaced by lines~\ref{alg:reducederror:2}--\ref{alg:reducederror:4}. In particular, line~\ref{alg:reducederror:2} computes the Top-$Q$ sparsification mask for $\vec g_k^t$, line~\ref{alg:reducederror:3} retrieves the sparsification mask of the incoming partial aggregate $\vec\gamma_{k+1}^t$, and line~\ref{alg:reducederror:4} sparsifies $\vec g_k^t$ after combining both sparsification masks.
\vspace{-10pt}
    \begin{algorithm}
	 \caption{Reduced-error sparse \cgls{ia} at node $k$}
	\label{alg:reducederror}
	\begin{algorithmic}[1]
		\StartIA{$\vec g_k^t$, $\vec\gamma_{k+1}^t$}
			\State Error feedback $\tilde{\vec g}_k^t \gets D_k \vec g_k^t + \vec e_k^{t-1}$ \label{alg:reducederror:1}
			\State Local sparsification mask $\vec m_k^t \gets s(\tilde{\vec g}_k^t, Q)$ \label{alg:reducederror:2}
			\State Incoming sparsification mask $\widetilde{\vec m}_{k+1}^t \gets \mathds 1(\gamma_{k+1}^t)$ \label{alg:reducederror:3}
			\State Sparsification $\bar{\vec g}_k^t \gets \mathds 1(\vec m_k^t + \widetilde{\vec m}_{k+1}^t) \circ \tilde{\vec g}_k^t$ \label{alg:reducederror:4}
			\State Update error $\vec e_k^t \gets \tilde{\vec g}_k^t - \bar{\vec g}_k^t$ \label{alg:reducederror:5}
			\State Incremental Aggregation $\vec\gamma_k^t \gets \bar{\vec g}_k^t + \vec\gamma_{k+1}^t$ \label{alg:reducederror:5}
		\EndIA{$\vec\gamma_k^t$}
	\end{algorithmic}
\end{algorithm} 
\vspace{-10pt}

This algorithm has the same communication cost as \cref{alg:vanilla} at a lower sparsification error. Since this directly corresponds to more information being transmitted to the \cgls{ps}, better training performance (in terms of accuracy) is expected. However, \cref{alg:reducederror} is neither optimal with respect to \cref{opt:mh} nor does it adhere to a strict communication budget.
\vspace{-5pt}
\subsection{Constant-Length Sparse Incremental Aggregation}
Equation \cref{opt:mh} also offers insight into the design of a sparse \cgls{ia} strategy that strictly adheres to a communication budget of $Q$ nonzero elements per hop. Indeed, the optimal sparsification strategy with respect to \cref{opt:mh} is $C(\vec\gamma_{k+1}^t, \tilde{\vec g}_k^t) = S(\tilde{\vec g}_k^t + \vec\gamma_{k+1}^t, \widetilde Q)$ and by setting $\widetilde Q = Q$, the outgoing partial aggregate $\vec \gamma_k^t = S(\tilde{\vec g}_k^t + \vec\gamma_{k+1}^t, Q)$ will have at most $Q$ nonzero entries. This immediately leads to \cref{alg:const}. Observe that the order of error feedback and \cgls{ia} in lines~\ref{alg:const:1} and~\ref{alg:const:2} is inconsequential, as long as the sparsification error in line~\ref{alg:const:4} is tracked correctly.
   \floatstyle{spaceruled}
\restylefloat{algorithm}
\begin{algorithm}
	\caption{Constant-length sparse \cgls{ia} at node $k$} \label{alg:const}
	\begin{algorithmic}[1]
		\StartIA{$\vec g_k^t$, $\vec\gamma_{k+1}^t$}
		\State Error feedback $\tilde{\vec g}_k^t \gets D_k \vec g_k^t + \vec e_k^{t-1}$ \label{alg:const:1}
		\State Incremental Aggregation $\tilde{\vec\gamma}_k^t \gets \tilde{\vec g}_k^t + \vec\gamma_{k+1}^t$  \label{alg:const:2}
		\State Sparsification $\vec\gamma_k^t \gets S(\tilde{\vec \gamma}_k^t, Q)$ \label{alg:const:3}
		\State Update error $\vec e_k^t \gets \tilde{\vec\gamma}_k^t - \vec\gamma_k^t$ \label{alg:const:4}
		\EndIA{$\vec\gamma_k^t$}
	\end{algorithmic}
\end{algorithm}
\floatstyle{ruled}
\restylefloat{algorithm}
\vspace{-5 pt}
\section{Time-Correlated Sparse IA} \label{sec:tcs}
In \cref{sec:algs}, we improved upon \cref{alg:vanilla} in two different ways. \Cref{alg:reducederror} has exactly the same communication cost as \cref{alg:vanilla}, but utilizes the bandwidth more efficiently. Instead, \cref{alg:const} fixes the issue of increasing communication cost with each hop and, thus, recovers the communication efficiency of \cgls{ia}. These strategies represent two extremes among potential improvements of \cref{alg:vanilla}:
we can either embrace the increasing communication cost and reduce the sparsification error, or we can strictly enforce the communications budget and obtain the superior communication performance of \cgls{ia}. A controllable trade-off between these two extremes can be obtained by combining sparse \cgls{ia} with \cgls{tcs}. The central idea of \cgls{tcs} \cite{Ozfatura2021a} is to compute a Top-$Q$ sparsification mask from the global model parameter vector. This leads to identical sparsification supports at all clients. 
Temporal dynamics in the model development are captured through a small number $Q_L$ of local additions to the global mask, which implements the desired trade-off.
\vspace{-5pt}
\subsection{Time-Correlated Sparse Incremental Aggregation}
The combination of \cgls{ia} and \cgls{tcs} operates on the same principles as \cref{alg:reducederror}. Let $Q_G$ and $Q_L$ denote the number of nonzero elements in the global and local sparsification masks, respectively. With $\vec w^t$ and $\vec w^{t-1}$ the current and previous iteration of the global model parameters, respectively, the effective gradient for the global model parameters is $\vec w^t - \vec w^{t-1}$. Then, the global sparsification mask is obtained from Top-$Q_G$ sparsification of $\vec w^t - \vec w^{t-1}$, i.e.,  $\vec m^t = s(\vec w^t - \vec w^{t-1}, Q_G)$.
The local sparsification mask $\vec m_k^t$ should capture the most relevant local gradient entries not yet part of the global mask. Thus, node $k$ computes $\vec m_k^t$ from the error-compensated gradient 
$\tilde{\vec g}_k^t$ after applying the global mask, i.e., $\vec m_k^t = s((\vec 1-\vec m^t)\circ \tilde{\vec g}_k^t, Q_L)$.
Similarly to \cref{alg:reducederror}, node $k$ can transmit further nonzero elements within the nonzero positions of $\vec\gamma_{k+1}^t$. Hence, the sparsified effective local gradient is $\bar{\vec g}_k^t = \mathds 1(\vec m^t + \vec m_k^t + \widetilde{\vec m}_{k+1}^t) \circ \tilde{\vec g}_k^t$ with $\widetilde{\vec m}_{k+1}^t = \mathds 1(\vec\gamma_{k+1}^t) - \vec m^t$.

Since the global mask $\vec m^t$ is known by all participants, it is not necessary to transmit the indices of these nonzero entries.
This leads to a considerable bandwidth reduction and a mixed storage format for the outgoing partial aggregate $\vec\gamma_k^t = \left[ \vec\Gamma_k^t,\enskip \vec\Lambda_k^t \right]$, where $\vec\Gamma_k^t$ contains the nonzero elements due to the global sparsification mask and $\vec\Lambda_k^t$ stores the elements due to local sparsification. Thus, the \cgls{ia} operation is
\begin{equation}\label{eq:tcsia}
	\begin{aligned}
		\vec\Gamma_k^t &= \vec\Gamma_{k+1}^t + \vec m^t \circ \bar{\vec g}_k^t
		\\
		\vec\Lambda^t &= \vec\Lambda_{k+1}^t + (\vec 1 - \vec m^t) \circ \bar{\vec g}_k^t
	\end{aligned}
\end{equation}
The complete procedure is stated in \cref{alg:tcs}, which extends \cref{alg:reducederror} with \cgls{tcs} in lines \ref{alg:tcs:1}, \ref{alg:tcs:2}, \ref{alg:tcs:3}, and \ref{alg:tcs:4}.

\vspace{-5pt}
   \begin{algorithm}
	\caption{Time-correlated sparse \cgls{ia} at node $k$}
	\label{alg:tcs}
	\begin{algorithmic}[1]
			\StartIA{$\vec g_k^t$, $\vec\gamma_{k+1}^t = \big[ \vec\Gamma_{k+1}^t,\enskip \vec\Lambda_{k+1}^t \big]$}
			\State Error feedback $\tilde{\vec g}_k^t \gets D_k \vec g_k^t + \vec e_k^{t-1}$
			\State Retrieve global mask $\vec m^t \gets s(\vec w^t - \vec w^{t-1}, Q_G)$ \label{alg:tcs:1}
			\State Local sparsification mask $\vec m_k^t \gets s((\vec 1-\vec m^t)\circ \tilde{\vec g}_k^t, Q_L)$ \label{alg:tcs:2}
			\State Incoming sparsification mask $\widetilde{\vec m}_{k+1}^t \gets \mathds 1(\gamma_{k+1}^t) - \vec m^t$
			\State Sparsification $\bar{\vec g}_k^t \gets \mathds 1(\vec m^t + \vec m_k^t + \widetilde{\vec m}_{k+1}^t) \circ \tilde{\vec g}_k^t$ \label{alg:tcs:3}
			\State Update error $\vec e_k^t \gets \tilde{\vec g}_k^t - \bar{\vec g}_k^t$
			\State Incremental Aggregation as in \cref{eq:tcsia} \label{alg:tcs:4}
		\EndIA{$\vec\gamma_k^t = \left[ \vec\Gamma_k^t,\enskip \vec\Lambda_k^t \right]$}
	\end{algorithmic}
\end{algorithm}
 
\vspace{-10pt}

\subsection{Constant-Length Time-Correlated Sparse IA}
The rationale behind communicating a small amount of locally selected nonzero elements is to allow new elements to enter the global mask during the training process.
This is also realized by the constant-length sparse \cgls{ia} procedure in \cref{alg:const}, which we combine with \cgls{tcs} in \cref{alg:tcsconst}. The vectors $\vec e_k^t$ and $\vec\Lambda_k^t$ are both implemented as $d$-dimensional sparse vectors. It is important to apply the error feedback in $\tilde{\vec g}_k^t$ (line~\ref{alg:tcsconst:1}) and not only in the computation of $\tilde{\vec\Lambda}_k^t$ (line~\ref{alg:tcsconst:3}), as the global mask might change between iterations.
    \begin{algorithm}
	\caption{Constant-length time-correlated sparse \cgls{ia}}
	\label{alg:tcsconst}
	\begin{algorithmic}[1]
			\StartIA{$\vec g_k^t$, $\vec\gamma_{k+1}^t = \big[ \vec\Gamma_{k+1}^t,\enskip \vec\Lambda_{k+1}^t \big]$}
			\State Error feedback $\tilde{\vec g}_k^t \gets D_k \vec g_k^t + \vec e_k^{t-1}$ \label{alg:tcsconst:1}
			\State Retrieve global mask $\vec m^t \gets s(\vec w^t - \vec w^{t-1}, Q_G)$ \label{alg:tcsconst:2}
			\State Incremental Aggregation $\begin{aligned}[t]
				\vec\Gamma_k^t &\gets \vec\Gamma_{k+1}^t + \vec m^t \circ \tilde{\vec g}_k^t \\
				\tilde{\vec\Lambda}_k^t &\gets \vec\Lambda_{k+1}^t + (\vec 1 - \vec m^t) \circ \tilde{\vec g}_k^t
			\end{aligned}$ \label{alg:tcsconst:3}
			\State Local sparsification $\vec\Lambda_k^t \gets S(\tilde{\vec\Lambda}_k^t, Q_L)$ \label{alg:tcsconst:4}
			\State Update error $\vec e_k^t \gets \tilde{\vec\Lambda}_k^t - \vec\Lambda_k^t$ \label{alg:tcsconst:5}

		\EndIA{$\vec\gamma_k^t = \left[ \vec\Gamma_k^t,\enskip \vec\Lambda_k^t \right]$}
	\end{algorithmic}
\end{algorithm}

\section{Communication Cost} \label{sec:cost}
The primary motivation for sparsification and \cgls{ia} is to reduce the communication cost of \cgls{fl}. The communication cost (in iteration $t$) of \cref{alg:vanilla,alg:reducederror,alg:const,alg:tcs,alg:tcsconst} directly depends on the number of transmitted nonzero elements $\sum_{k=1}^K \Vert \vec\gamma_k^t \Vert_0$ and the storage representation of each element. For the time-correlated sparse \cgls{ia} method in \cref{alg:tcs}, the number of outgoing nonzero elements at node $k$ is $\Vert\vec\gamma_k^t\Vert_0 = \Vert\vec\Gamma_k^t\Vert_0 + \Vert\vec\Lambda_k^t\Vert_0$. While $\Vert\vec\Gamma_k^t\Vert_0$ is deterministically $Q_G$, $\Vert\vec\Lambda_k^t\Vert_0$ is a random number. Each nonzero element in $\vec\Lambda_k^t$ requires $\omega$ bit for the numerical value and additional $\lceil\log_2 d\rceil$ bit to store its location in $\vec\gamma_k^t$, while each element in $\vec\Gamma_k^t$ only requires a total of $\omega$ bit. Thus, the expected total communication cost for \cref{alg:tcs} is
\begin{multline} \label{eq:comcost}
	\sum\nolimits_{k=1}^K \left( \omega \mathds E\left[ \Vert\vec\Gamma_k^t\Vert_0 \right] + \left( \omega + \lceil\log_2 d\rceil \right)\mathds E\left[ \Vert\vec\Lambda_k^t\Vert_0 \right] \right)
	\\
	=
	K \omega Q_G + \left( \omega + \lceil\log_2 d\rceil \right) \sum\nolimits_{k=1}^K \mathds E\left[ \Vert\vec\Lambda_k^t\Vert_0 \right].
\end{multline}
Exact analytical expressions for $\mathds E\left[ \Vert\vec\Lambda_k^t\Vert_0 \right]$ are challenging to obtain. However, the following upper bound can be derived along the lines of \cite[Prop.~1]{Razmi2024} by observing that $\vec\Lambda_k^t$ is effectively a vector of dimension $d - Q_G$. The proof is omitted due to space limitations.
\begin{proposition} \label{prop:ceff}
	The expected number of nonzero elements due to local sparsification communicated by \cref{alg:tcs} over $K$ hops is upper bounded as
	\begin{multline}
		\textstyle
		\sum_{k=1}^K \mathds E\left[ \Vert\vec\Lambda_k^t\Vert_0 \right] \le \left( d-Q_G \right)\\
		\textstyle
			 \Big(K + 1 - \frac{d-Q_G}{Q_L} \Big( 1 - \Big(1-\frac{Q_L}{d-Q_G}\Big)^{K+1} \Big)\Big)
	\end{multline}
	if $Q_L > 0$ and zero otherwise.
\end{proposition}

From a communications cost perspective, \cref{alg:vanilla,alg:reducederror} are equivalent to \cref{alg:tcs} with $Q_G = 0$ and $Q_L = Q$. Furthermore, the communications cost of \cref{alg:tcsconst} follows from \cref{eq:comcost} by observing that $\mathds E\left[ \Vert\vec\Lambda_k^t\Vert_0 \right] = Q_L$ as
$K \omega Q_G + \left( \omega + \lceil\log_2 d\rceil \right) K Q_L$.
Finally, since
\cref{alg:const} might be regarded as a special case of \cref{alg:tcsconst} with $Q_G = 0$ and $Q_L = Q$, its communication cost is
$K Q \left( \omega +  \lceil\log_2 d\rceil \right)$.

\begin{figure*}
	\centering
	\subfloat[Absolute]{%
	\begin{tikzpicture}
		\begin{axis} [
				thick,
				font = {\small},
				xlabel={Number of clients $K$},
				ylabel={Total transmitted data\\{}[\si{\mega\bit}/iteration]},
				ylabel near ticks,
				ylabel style = {align=center},
				grid=major,
				minor x tick num = 1,
				minor y tick num = 1,
				legend entries = {CL-TC-SIA, TC-SIA, CL-SIA, RE-SIA, SIA},
				no markers,
				legend cell align=left,
				legend pos = {north west},
				scaled y ticks = base 10:-6,
				ytick scale label code/.code={},
				legend style = {font={\footnotesize}},
				width=\axisdefaultwidth,
				height=.65*\axisdefaultheight,
			]

			\addplot[blue, dashed] table [x=Total clients, y=commCost, col sep=comma] {src/CommCost_algo_1.csv};
			\addplot[blue] table [x=Total clients, y=commCost, col sep=comma] {src/CommCost_algo_2.csv};
			\addplot[black!40!green, dashed] table [x=Total clients, y=commCost, col sep=comma] {src/CommCost_algo_4.csv};
			\addplot[black] table [x=Total clients, y=commCost, col sep=comma] {src/CommCost_algo_5.csv};
			\addplot[black!40!green] table [x=Total clients, y=commCost, col sep=comma] {src/CommCost_algo_6.csv};
		\end{axis}
	\end{tikzpicture}%
	\label{fig:comeff}%
	}
	\hfill%
	\subfloat[Relative]{%
	\begin{tikzpicture}
		\begin{axis} [
				thick,
				font = {\small},
				xlabel={ Number of clients $K$},
				ylabel={Total transmitted data\\{}[Normalized]},
				ylabel near ticks,
				yticklabel pos = right,
				ylabel style = {align=center},
				grid=major,
				minor x tick num = 1,
				minor y tick num = 1,
				legend entries = { Unicast, IA,CL-TC-SIA, TC-SIA,CL-SIA, RE-SIA, SIA},
				no markers,
				legend cell align=left,
				legend columns = 2,
				legend pos = {north west},
				legend style = {font={\footnotesize}},
				width=\axisdefaultwidth,
				height=.65*\axisdefaultheight,
			]

			\pgfplotstableread[col sep=comma]{src/noSparse_noIA.csv}\tbl  
			\addplot[gray, dashed] table [x=Total clients, y=data_noIA_noSparse_normalized] {\tbl};
			\addplot[gray] table [x=Total clients, y=data_IA_noSparse_normalized] {\tbl};

			\addplot[blue, dashed] table [x=Total clients, y=commCost, col sep=comma] {src/CommCostNorm_algo_1.csv};
			\addplot[blue] table [x=Total clients, y=commCost, col sep=comma] {src/CommCostNorm_algo_2.csv};
			\addplot[black!40!green, dashed] table [x=Total clients, y=commCost, col sep=comma] {src/CommCostNorm_algo_4.csv};
			\addplot[black] table [x=Total clients, y=commCost, col sep=comma] {src/CommCostNorm_algo_5.csv};
			\addplot[black!40!green] table [x=Total clients, y=commCost, col sep=comma] {src/CommCostNorm_algo_6.csv};
		\end{axis}
	\end{tikzpicture}%
	\label{fig:comeff_norm}%
	}
	\caption{Total transmitted data per global iteration for fixed $Q = 78$ with respect to the number of clients.}
	\vspace{-3mm}
\end{figure*}
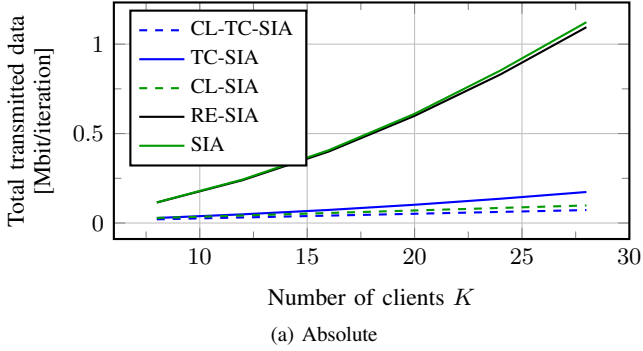
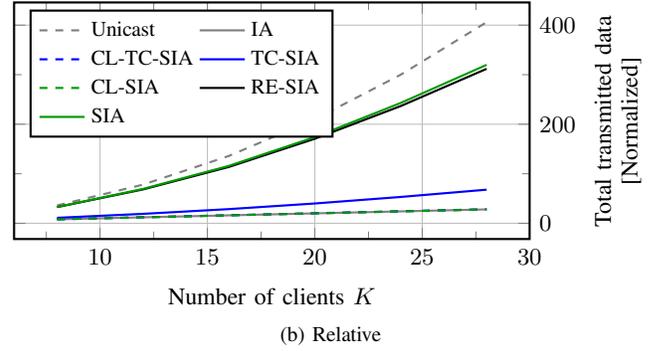

\section{Numerical Evaluation} \label{sec:numeval}
We evaluate the performance of the proposed sparse \cgls{ia} methods for a logistic regression model with $d = 7850$ trainable parameters on the MNIST data set \cite{MNIST}. This model is trained using \cgls{sgd} with batch size 20 and learning rate 0.1.
The first experiment uses a fixed $Q =  78$, corresponding to retaining \SI{1}{\percent} of nonzero elements. Following \cite{Ozfatura2021a}, we set $Q_L = 8$ in \cref{alg:tcs,alg:tcsconst} to \SI{10}{\percent} of $Q$, and $Q_G = Q - Q_L$. We refer to \cref{alg:vanilla,alg:reducederror,alg:const,alg:tcs,alg:tcsconst} as SIA, RE-SIA, CL-SIA, TC-SIA, and CL-TC-SIA, respectively, where CL and TC refer to the constant-length and time-correlated properties, respectively.

\Cref{fig:comeff} shows the total data transmitted in the aggregation step of a single iteration, averaged over the complete training process. As expected, the communication cost for SIA and RE-SIA is significantly higher than for the other algorithms. The TC-SIA approach shows the same quadratic growth, but at a much reduced rate. Finally, the CL algorithms are transmitting the smallest amount of data. The constant gap between these two is due to \cgls{tcs} transmitting $Q_G$ fewer indices.

The communication efficiency of \cgls{ia}, adjusted to exclude any sparsification effects, is displayed in \cref{fig:comeff_norm}. There, we normalized the total transmitted data of each algorithm with the size of a single gradient transmission. We also include the normalized communication costs for the case without any \cgls{ia}, i.e., conventional routing with multiple unicast transmissions, and \cgls{ia} without any sparsification. Most notably, we observe that CL-SIA and CL-TC-SIA are meeting the same performance as \cgls{ia} without sparsification. This implies that those two algorithms do not suffer from the decreasing efficiency of \cgls{ia} under increasing sparsification ratios. We also see that, while not being able to fully utilize the benefits of \cgls{ia}, SIA and RE-SIA still show much better performance than conventional routing.

\begin{figure}
	\centering
	\begin{tikzpicture}
		\begin{axis} [
				thick,
				font = {\small},
				xlabel={Iteration},
				ylabel={Test Accuracy},
				ylabel near ticks,
				grid=major,
				ymin = 0.2,
				ymax = 0.92,
				minor x tick num = 1,
				minor y tick num = 1,
				legend entries = {CL-TC-SIA, TC-SIA, CL-SIA, RE-SIA, SIA},
				no markers,
				legend cell align=left,
				legend pos = {south east},
				legend style = {font={\footnotesize}},
				width=1*\axisdefaultwidth,
				height=.65*\axisdefaultheight,
			]

			\addplot[blue, dashed] table [x=Round, y=commCost, col sep=comma] {src/Accu_algo_1.csv};
			\addplot[blue] table [x=Round, y=commCost, col sep=comma] {src/Accu_algo_2.csv};
			\addplot[black!40!green, dashed] table [x=Round, y=commCost, col sep=comma] {src/Accu_algo_4.csv};
			\addplot[black] table [x=Round, y=commCost, col sep=comma] {src/Accu_algo_5.csv};
			\addplot[black!40!green] table [x=Round, y=commCost, col sep=comma] {src/Accu_algo_6.csv};
		\end{axis}
	\end{tikzpicture}
	\caption{Test accuracy for a fixed $Q = 78$ and $K = 28$ clients.}
	\label{fig:accuracy}
\end{figure}
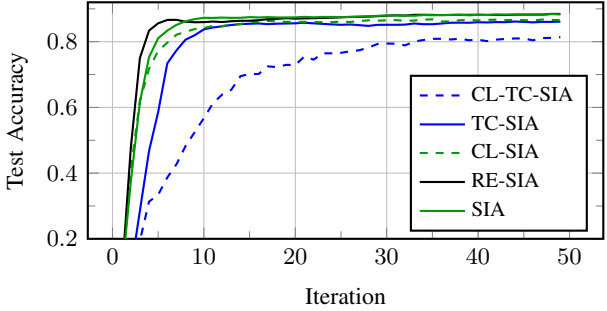

\Cref{fig:accuracy} shows convergence of the training process in terms of the test accuracy for $K = 28$ clients.
The higher communication cost of SIA and RE-SIA directly translates to better learning performance. This is not surprising, as these algorithms send much more information to the \cgls{ps}. However, the reduced sparsification error of RE-SIA only results in a slight edge in convergence speed over SIA. Moreover, convergence of CL-SIA and TC-SIA is only slightly worse than SIA despite their much lower communication cost. We further observe that the convergence speed of CL-TC-SIA is severely impaired. This is likely due to the much smaller effective $Q_L$, resulting in slower adaption to temporal dynamics.

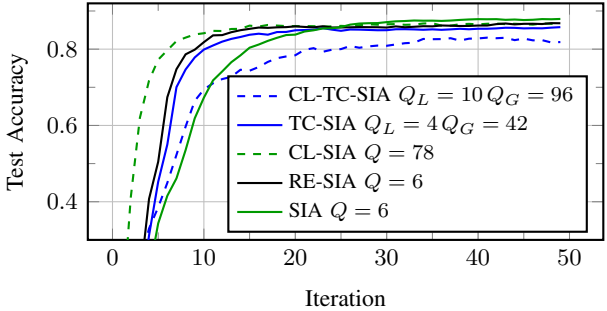
\begin{figure}
	\centering
	\begin{tikzpicture}
		\begin{axis} [
				thick,
				font = {\small},
				xlabel={Iteration},
				ylabel={Test Accuracy},
				ylabel near ticks,
				ymin = 0.3,
				ymax = 0.92,
				grid=major,
				minor x tick num = 1,
				minor y tick num = 1,
				legend entries = {CL-TC-SIA $Q_L = 10 \, Q_G = 96$, TC-SIA $Q_L = 4 \,Q_G = 42$, CL-SIA $Q = 78$, RE-SIA $Q = 6$, SIA $Q = 6$},
				no markers,
				legend cell align=left,
				legend pos = {south east},
				legend style = {font={\footnotesize}},
				width=1*\axisdefaultwidth,
				height=.65*\axisdefaultheight,
			]

			\addplot[blue, dashed] table [x=Round, y=commCost, col sep=comma] {src/AccuCal_algo_3.csv};
			\addplot[blue] table [x=Round, y=commCost, col sep=comma] {src/AccuCal_algo_4.csv};
			\addplot[black!40!green, dashed] table [x=Round, y=commCost, col sep=comma] {src/AccuCal_algo_5.csv};
			\addplot[black] table [x=Round, y=commCost, col sep=comma] {src/AccuCal_algo_1.csv};
			\addplot[black!40!green] table [x=Round, y=commCost, col sep=comma] {src/AccuCal_algo_2.csv};
		\end{axis}
	\end{tikzpicture}
	\caption{Test accuracy for $K = 28$ clients under (approximately) equal average bandwidth usage of \SI{98}{\kilo\bit} per global iteration.}
	\label{fig:accuracy_calibrated}
\end{figure}
For a clearer picture of the bandwidth-efficiency, \cref{fig:accuracy_calibrated} shows the test accuracy under approximately equal total bandwidth usage. To this end, we consider the same scenario as in \cref{fig:accuracy} but varied $Q$ such that each algorithm transmits the same amount of data as CL-SIA. For TC-SIA and CL-TC-SIA, we maintain the split $Q_L = 0.1 Q$ and $Q_G = 0.9 Q$. The result is a 
slightly higher bandwidth usage for CL-TC-SIA and significantly less for SIA, RE-SIA, and TC-SIA. In terms of training, we see that CL-SIA, RE-SIA, and TC-SIA converge much faster than SIA, with CL-SIA having best performance.

\section{Conclusions}
We have considered efficient communication for \cgls{fl} in \cgls{mh} networks. The \cgls{soa} approach for collecting intermediate \cgls{fl} results in such a system is \cgls{ia} \cite{Razmi2022,Chen2022b}, which shows diminishing returns under gradient sparsification \cite{Razmi2024}. We have developed four novel algorithms to address this issue and evaluated their performance numerically. These results show a distinct advantage of \cref{alg:const} over all other methods, both in terms of communication cost and \cgls{fl} performance. However, our results also show a minor gap in final test accuracy over the \cgls{soa}. This will be subject to further investigation in future work, starting with a rigorous convergence analysis of \cref{alg:const}.

\bibliography{IEEEtrancfg,IEEEabrv,spawc}

\begin{thebibliography}{10}
\providecommand{\url}[1]{#1}
\csname url@samestyle\endcsname
\providecommand{\newblock}{\relax}
\providecommand{\bibinfo}[2]{#2}
\providecommand{\BIBentrySTDinterwordspacing}{\spaceskip=0pt\relax}
\providecommand{\BIBentryALTinterwordstretchfactor}{4}
\providecommand{\BIBentryALTinterwordspacing}{\spaceskip=\fontdimen2\font plus
\BIBentryALTinterwordstretchfactor\fontdimen3\font minus \fontdimen4\font\relax}
\providecommand{\BIBforeignlanguage}[2]{{%
\expandafter\ifx\csname l@#1\endcsname\relax
\typeout{** WARNING: IEEEtran.bst: No hyphenation pattern has been}%
\typeout{** loaded for the language `#1'. Using the pattern for}%
\typeout{** the default language instead.}%
\else
\language=\csname l@#1\endcsname
\fi
#2}}
\providecommand{\BIBdecl}{\relax}
\BIBdecl

\bibitem{Razmi2024}
N.~Razmi, B.~Matthiesen, A.~Dekorsy, and P.~Popovski, ``On-board federated learning for satellite clusters with inter-satellite links,'' \emph{{IEEE} Trans. Commun.}, Jan. 2024.

\bibitem{mcmahan2017communication}
B.~McMahan, E.~Moore, D.~Ramage, S.~Hampson, and B.~Ag{\"u}era~y Arcas, ``Communication-efficient learning of deep networks from decentralized data,'' in \emph{Artif. Intell. Statist. (AISTATS)}, Fort Lauderdale, FL, Apr. 2017.

\bibitem{bengio2016}
I.~Goodfellow, Y.~Bengio, and A.~Courville, \emph{Deep Learning}.\hskip 1em plus 0.5em minus 0.4em\relax Cambridge, MA, USA: MIT Press, 2016.

\bibitem{network2022}
B.~Matthiesen, N.~Razmi, I.~Leyva-Mayorga, A.~Dekorsy, and P.~Popovski, ``Federated learning in satellite constellations,'' \emph{{IEEE} Netw.}, May 2023.

\bibitem{Razmi2022}
N.~Razmi, B.~Matthiesen, A.~Dekorsy, and P.~Popovski, ``Ground-assisted federated learning in {LEO} satellite constellations,'' \emph{{IEEE} Wireless Commun. Lett.}, vol.~11, no.~4, pp. 717--721, Apr. 2022.

\bibitem{Razmi2022a}
------, ``On-board federated learning for dense {LEO} constellations,'' in \emph{IEEE Int. Conf. Commun.}, Seoul, Korea, May 2022.

\bibitem{Chen2022b}
X.~Chen, G.~Zhu, Y.~Deng, and Y.~Fang, ``Federated learning over multihop wireless networks with in-network aggregation,'' \emph{{IEEE} Trans. Wireless Commun.}, vol.~21, no.~6, pp. 4622--4634, Jun. 2022.

\bibitem{Aji2017}
A.~F. Aji and K.~Heafield, ``Sparse communication for distributed gradient descent,'' in \emph{Conf. Empir. Methods Nat. Lang. Process.}, Sep. 2017.

\bibitem{Alistarh2018}
D.~Alistarh \emph{et~al.}, ``The convergence of sparsified gradient methods,'' in \emph{Adv. Neural Inf. Process. Syst.}, vol.~31, 2018.

\bibitem{Ozfatura2021a}
\BIBentryALTinterwordspacing
E.~Ozfatura, K.~Ozfatura, and D.~Gündüz, ``Time-correlated sparsification for communication-efficient federated learning,'' Jan. 2021. [Online]. Available: \url{http://arxiv.org/abs/2101.08837}
\BIBentrySTDinterwordspacing

\bibitem{Sahu2021}
A.~Sahu \emph{et~al.}, ``Rethinking gradient sparsification as total error minimization,'' in \emph{Adv. Neural Inf. Process. Syst.}, vol.~34, 2021.

\bibitem{MNIST}
Y.~{LeCun}, C.~Cortes, and C.~J.~C. Burges. The {MNIST} database of handwritten digits.

\end{thebibliography}
\end{document}